\documentclass[fleqn,twoside]{article}
\usepackage{espcrc2}
\usepackage{amsmath}

\usepackage{graphicx}
\renewcommand{\vec}{\mathbf}
\renewcommand{\Im}{\text{Im}}
\title{Temperature-dependent electronic structure and magnetic stability of
thin ferromagnetic films}

\author{W.~Nolting\address{Lehrstuhl Festk{\"o}rpertheorie,
Institut f{\"u}r Physik, Humboldt-Universit{\"a}t zu Berlin,\\ 
Invalidenstr.~110, 10115 Berlin, Germany} and C.~Santos\addressmark}
\begin{document}
\begin{abstract}
We study correlation effects and temperature dependencies in the electronic structure of thin  
ferromagnetic local-moment films. In a first step the Kondo-lattice model is investigated as a  
candidate for a proper representation of local-moment ferromagnets. Magnetic and electronic  
key-quantities as the Curie-temperature and the quasiparticle density of states are derived with  
previously tested many-body procedures. It is shown that the magnetic properties can be
interpreted exclusively in terms of the temperature-dependent electronic quasiparticle structure.  
An extended RKKY theory leads to effective Heisenberg exchange integrals, which turn out  
to be functionals of the conduction electron selfenergy, getting therewith a remarkable
temperature and band occupation dependence. 
 
In a second step the model studies are combined with tight binding-LMTO bandstructure
calculations in order to get for real ferromagnetic films quasiparticle densities of states and
quasiparticle bandstructures. The proposed method avoids the double-counting of relevant
interactions and takes into account the correct symmetry of the atomic orbitals. Special results are  
given for thin ferromagnetic EuO (100) films. The Curie temperature
$T_{\textrm{C}}$ of the EuO film  
turns out to be strongly thickness-dependent, starting from a very low value ($\simeq 15K$) for the  
monolayer and reaching the bulk value at about 30 layers. For a 20-layer film we predict the  
existence of a surface state, the temperature-behaviour of which can lead to a surface
halfmetal-insulator transition. 
\end{abstract}
\maketitle
\section{Introduction}\label{sec:intro}
All key-quantities of ferromagnetism (Curie-temperature $T_{\textrm{C}}$,
magnetic moment $\mu$, magnetization $M(T)$, susceptibility $\chi$,\dots) are in the last
analysis consequences of the electronic structure  
of the respective magnetic material. To understand ferromagnetism therefore means to
understand the temperature-dependent electronic structure. Our method, which we use to get
reliable information in this respect, consists of three steps. First we choose a proper
theoretical model, defined by its Hamiltonian 
\begin{equation}\label{eq:hamil} 
  H=H_{0}+H_{I}
\end{equation}
more strictly by the interaction part $H_{I}$ which shall incorporate
all those interactions  which are considered as relevant for the
physical  problem under study. The single-particle part, on the  
other hand, shall cover, besides the usual kinetic energy and the
periodic  lattice potential, the influences of all the other
interactions  which are not directly accounted for by $H_I$. By definition
they are not decisive for magnetism and the characteristic
temperature-dependent electronic structure, but nevertheless, they may
determine the rough structure of the energy spectrum being therefore
non-negligible  if our study aims at a more or less quantitative electronic  
structure description. We therefore perform in the second step a first
principles  bandstructure calculation on the basis of density functional
theory  (DFT) and use the results as single-particle energies in
$H_{0}$. So  we guarantee that all the other interactions show up in
$H_0$  in an averaged but fairly realistic manner. However, one has
carefully  to avoid a double counting of  
just the relevant interactions, once explicitly in $H_{I}$ and then once
more implicitly in the renormalized single-particle energies. How to
circumvent this serious and well-known double  
counting problem shall be explained for the actual problem at a later
stage of this paper. 
 
As a third step a many-body formalism is used for (\ref{eq:hamil}) to
work out  how the effective single-particle energies change under the
relevant  interactions $H_I$ into a temperature- and
concentration-dependent  selfenergy $\Sigma_{\vec{k}\sigma}(E)$ from which we derive the
quasiparticle  bandstructure (Q-BS). The latter directly corresponds to
the  data of an angle- and spin-resolved photoemission experiment. 
 
We proceed in the following exactly along the line given by the just-developed concept. To  
do the first step, the fixing of a proper theoretical model, requires of course above all to agree  
upon which type of magnetic material shall be studied. We concentrate ourselves in this paper  
on so-called local-moment systems, i.~e.~on materials, the magnetic properties of which are  
due to strictly localized moments, while the conductivity properties are provoked by extended  
band states. Typical examples are magnetic insulators such as EuO, EuS \cite{WAC79} and magnetic  
metals as Gd \cite{DDN98}, which all have strictly localized moments
because of the half-filled $4f$ shell  
of the rare earth ion (Eu$^{2+}$, Gd$^{3+}$). Many striking features of
these materials  can be traced back  
to an intimate correlation between the localized magnetic moments and the extended band  
states. The same interaction is considered responsible for the
properties of the  intensively investigated diluted magnetic semiconductors like
Ga$_{1-x}$Mn$_{x}$As. The ion creates simultaneously a localized 
$S=5/2$ moment and an  itinerant hole in the GaAs valence band \cite{OHN98}, the  
interaction of which leads to ferromagnetism already for very low
concentrations  $x$. Another burning issue in this respect are the
colossal magnetoresistance (CMR) materials \cite{RAM97} such as  
La$_{1-x}$(Ca, Sr)$_x$MnO$_3$ which exhibit a rich magnetic phase diagram
as function of $x$. For $0.2 \le x \le 0.4$ the original insulating
antiferromagnetic parent compound LaMnO$_3$ becomes a ferromagnetic
metal. This is ascribed to a homogeneous valence mixing
(Mn$^{3+}_{1-x}$Mn$^{4+}_x$).  The three  
$5d-t_{2g}$ electrons in Mn$^{3+}$ form a localized $S = 3/2$ moment
while  the additional $5d-e_g$ electron  
is itinerant. Again the local moment-itinerant electron correlation is likely responsible for  
many typical features of the CMR materials.
\section{Kondo Lattice Model (KLM)}\label{sec:KLM} 
A model which provides an at least qualitatively correct insight into the physics of the local- 
moment systems is the Kondo-lattice model, which shall now be investigated in detail. 
\subsection{Model Hamiltonian, Exact Limiting Case}\label{ssec:limcase}
The KLM describes interacting local moments (spins $\vec{S}_i$) and itinerant
electrons in a  nondegenerate $s$-band:
\begin{equation}\label{eq:KLM}
H=\sum_{ij\sigma}T_{ij}c^{\dagger}_{i\sigma}c^{}_{j\sigma}-J\sum_{j}
\boldsymbol{\sigma}_{j}\cdot\vec{S}_{j}
\end{equation}
$c^{\dagger}_{i\sigma}(c^{}_{i\sigma})$ creates (annihilates) a band
electron with  spin $\sigma$ ($\sigma = \uparrow, \downarrow$) at the
lattice  site $\vec{R}_i$. $T_{ij}$ are the hopping integrals. The 
''a priori'' uncorrelated electrons are exchange coupled to the local  
moments, where this coupling is considered an on-site interaction of the
electron spin $\boldsymbol{\sigma}_i$ and  
the localized spin $\vec{S}_i$. $J$ is the coupling constant. Using second
quantization for the electron spin operator the interaction term reads: 
\begin{eqnarray}\label{eq:Hsf}
 \lefteqn{H_{I}=-\frac{J}{2}\sum_{j}\left(S^z_j(n^{}_{j\uparrow}-n^{}_{j\downarrow})
  +S^+_jc^{\dagger}_{j\downarrow}c^{}_{j\uparrow}+\right.}\nonumber\\
  &&\left.\hspace{4.5em}+S^-_jc^{\dagger}_{j\uparrow}c^{}_{j\downarrow}\right)
\end{eqnarray} 
($n_{j\sigma}=c^{\dagger}_{j\sigma}c^{}_{j\sigma}$). The first term
represents an ''Ising-type'' interaction while the two others refer  
to spin exchange processes. The latter give rise to some of the most
typical  KLM properties.  
Spin exchange may happen in three different elementary processes: magnon
emission by a $\downarrow$ electron, magnon absorption by a
$\uparrow$ electron and formation of a
quasiparticle known as ''magnetic polaron''. The quasiparticle can be
understood as a propagating electron dressed by a  
virtual cloud of magnons corresponding to a polarization of the immediate localized spin  
neighbourhood. 
 
\begin{figure}[ht]
\includegraphics[width=0.475\textwidth]{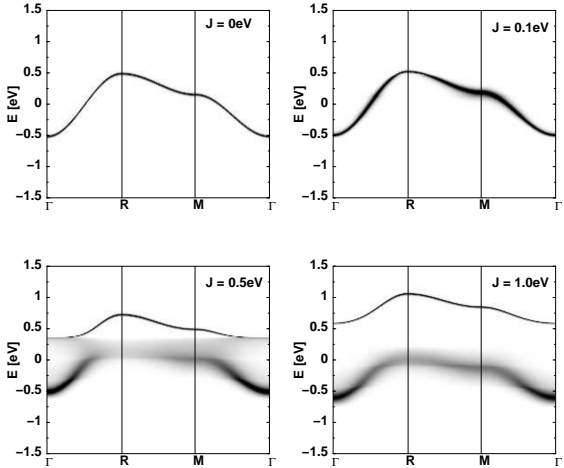}
\caption{Down-spin quasiparticle bandstructure of a ferromagnetically
  saturated semiconductor along several symmetry directions for
  different exchange couplings $J$. The degree of blackening in  
  the bandstructure is a measure of the magnitude of the spectral
  density. Parameters: $S = 1/2$, $W = 1eV$, s.c.~lattice. 
  $J = 0$ means the interaction-free Bloch-dispersion.}
\label{fig:QDOS}
\end{figure}

The sophisticated many-body problem provoked by the KLM-Hamiltonian (\ref{eq:KLM}) can in general  
not be solved exactly. Luckily there exists a non-trivial limiting case which is rigorously  
treatable exhibiting all the just-mentioned important elementary exchange processes. It is the  
situation of a ferromagnetically saturated semiconductor (EuO at $T = 0$), i.~e.~a single electron  
in an otherwise empty conduction band coupled to a saturated moment system. In this case the  
$\uparrow$ spectrum is extremely simple because the $\uparrow$ electron
has no chance to exchange its spin  
with the parallel aligned spin system. Only the Ising-type interaction
in (\ref{eq:Hsf}) takes  care for a  
rigid shift of the spectrum of $-\frac{1}{2}JS$. The quasiparticle
density of states (Q-DOS)  is identical  
to the free Bloch density of states (B-DOS),
$\rho_{\uparrow}(E)=\rho_0\left(E+\frac{1}{2}JS\right)$, expect for the trivial shift  
of $-\frac{1}{2}JS$. On the contrary, real correlation effects make
the $\downarrow$ spectrum highly non-trivial.  
Fig.~\ref{fig:QDOS} shows the $\downarrow$-Q-BS  
as density plot of the respective spectral density. The degree of blackening is a measure for  
the height of the quasiparticle peak. Rather moderate effective exchange couplings $JS/W$ are  
already sufficient to split the energy dispersion into two branches. The sharp high-energy one  
belongs to the magnetic polaron, which has in this case even an infinite
lifetime (''bound state''). The low-energy branch is stronger washed
out. It is due to magnon emission by the $\downarrow$  
electron which thereby reverses its spin. The magnon can carry away any
wave  vector from the first Brillouin zone. The spectrum (''scattering spectrum'') is
therefore  in general rather  
broad. Because of the spinflip it extends over just that energy region
where $\rho_{\uparrow}(E)\ne 0$. Surprisingly, however, the broad
scattering part is often bunched together to a rather prominent  
quasiparticle dispersion. The splitting of the original Bloch dispersion into two quasiparticle  
branches is a typical correlation effect, by no means reproducible by a single-particle theory. 
\subsection{Many-Body Evaluation}\label{ssec:eval}
For the general case (finite temperature, finite band occupations) the many-body problem of  
the KLM can only be solved approximately. A common feature of many approaches is the  
following structure of the conduction electron selfenergy \cite{NRM97,NRR01} 
\begin{equation}\label{eq:selfenergy} 
\Sigma_{\vec{k}\sigma}(E)=-\frac{1}{2}Jz_{\sigma}
\langle S^z\rangle+J^2\:D_{\vec{k}\sigma}(E)                    
\end{equation} 
If restricting to the first term, only, one has the mean-field approach
to  the KLM, which is correct for sufficiently weak couplings $J$. This
part is mainly due to the Ising-interaction in (\ref{eq:Hsf}). Without
the second term, it would lead to a spin-polarized splitting of the conduction  
band ($z_{\sigma}=\delta_{\sigma\uparrow}-\delta_{\sigma\downarrow}$). The more complicated
second part contains the spin exchange processes  
and the polaron formation. In a previous paper \cite{NRM97} we have proposed a moment conserving  
decoupling approach (MCDA) for a set of properly defined Green functions that correctly  
reproduces the exactly solvable limiting cases of the KLM. For details of the derivation the  
reader is referred to ref.~\cite{NRM97}. The MCDA demonstrates that the selfenergy
carries a distinct temperature-dependence which is brought into play by
two different types of spin correlations.  
There are mixed correlations such as $\langle
S^z_in^{}_{i\sigma}\rangle$,
$\langle S^+_ic^{\dagger}_{i\downarrow}c^{}_{i\uparrow}\rangle$,\dots built up by combinations of
localized-spin and itinerant-electron operators. Luckily all these correlations can be expressed  
via the spectral theorem by one of the Green functions involved in the MCDA. There is no  
need for further approximations. The second type of correlations are
pure local-moment correlations: $\langle S^z_i\rangle$, 
$\langle S^{\pm}_iS^{\mp}_i\rangle$,$\langle (S^z_i)^3\rangle$,\dots which need a special treatment. 
 
We use a ''modified'' RKKY (M-RKKY) theory \cite{NRM97,SNO01} which exploits a
mapping of the interband exchange (\ref{eq:Hsf}) to an effective Heisenberg model, 
\begin{equation}\label{eq:Hf}
H_f=-\sum_{ij}J_{ij}\:\vec{S}_{i}\cdot\vec{S}_j,
\end{equation}                  
by averaging out the conduction electron degrees of freedom: 
\begin{equation}\label{eq:Hi}
H_{I}\longrightarrow\langle
H_{I}\rangle^{(c)}=-J\sum_{j}\vec{S}_j\langle\boldsymbol{\sigma}_j\rangle^{(c)}
\longrightarrow H_f
\end{equation}
$\langle\dots\rangle^{(c)}$ means averaging in the subspace of the conduction
electrons. Details of the applied  
Green-function procedure can be found in refs.~\cite{NRM97,SNO01}. The
result are  effective exchange integrals $J_{ij}$ in (\ref{eq:Hf}): 
\begin{eqnarray}\label{eq:Jij}
\lefteqn{J_{ij}=\frac{J^2}{4\pi N^2}\sum_{\vec{k},\vec{q},\sigma}e^{i\vec{q}
\cdot(\vec{R}_i-\vec{R}_j)}\int\limits_{-\infty}^{+\infty}dE\:f_{-}(E)\times}\\\nonumber
&&\hspace{-3ex}\Im\left[\left(E-\varepsilon_{\vec{k}}+i0^+\right)
\left(E-\varepsilon_{\vec{k}+\vec{q}}-\Sigma_{\vec{k}+\vec{q}\sigma}(E)\right)\right]^{-1}
\end{eqnarray}                
$f_-(E)$ is the Fermi function. Most important is the appearance of the
conduction electron selfenergy on the right hand side. Neglecting $\Sigma_{\sigma}$ leads to
the ''conventional'' RKKY formula with  
$J_{ij}\sim J^2$, as a result of second order perturbation theory. By $\Sigma_{\sigma}$
higher order  conduction electron  
spin polarization terms enter the M-RKKY as well as a distinct temperature-dependence. With  
(\ref{eq:Hf}) and (\ref{eq:Jij}) we calculate the above-mentioned local
moment correlations, applying a standard  
Tyablikow-approximation to the Heisenberg model \cite{SNO01}. Together
with (\ref{eq:selfenergy}) a closed system of  
equations is built up which can be solved self-consistently for all electronic or magnetic  
KLM-properties, we are interested in. Results are presented in the next
Section. 
\subsection{KLM-Bulk Properties}\label{ssec:KLM-BULK}
\begin{figure}[ht]
\includegraphics[width=0.485\textwidth]{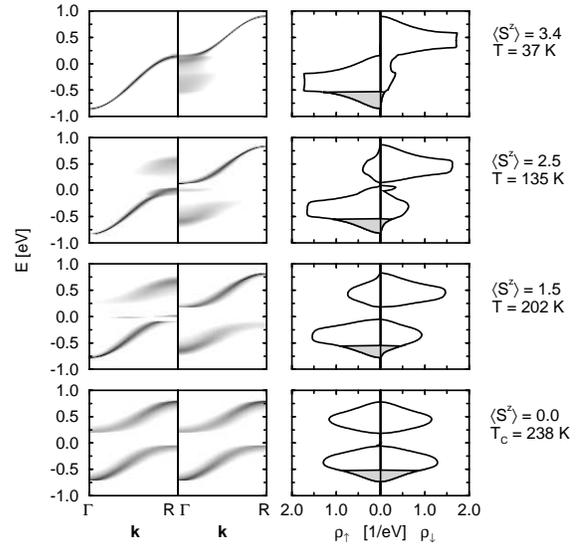}
\caption{Quasiparticle bandstructure as a function of wave-vector
  (left column) and quasiparticle density of states as a function of
  energy (right column) for four different temperatures calculated  
  within the MCDA \cite{NRM97}. 
 Parameters: $S = 7/2$, $W = 1eV$, $J = 0.2 eV$, $n = 0.2$, s.c.~lattice.}
\label{fig:QDOSa}
\end{figure}

Fig.~\ref{fig:QDOSa} shows the Q-BS and the Q-DOS for a model system on
a s.c.~lattice with $S = 7/2$, a  
moderate exchange coupling $J = 0.2eV$ and a band occupation $n = 0.2$. For this parameter  
constellation the self-consistently calculated Curie temperature amounts to $238K$. At low  
temperatures ($T = 37K$) the local moments are almost saturated, and
therefore  the $\uparrow$-dispersion and the Q-DOS $\rho_{\uparrow}$ are
practically identical to
the respective ''free''functions according to the limiting case discussed
in Sect.~\ref{ssec:eval}. The $\downarrow$ spectrum, however, is more
complicated. In the Q-BS
the scattering states are clearly visible near the \textrm{$\Gamma$}
point, while near the R  
point the magnetic polaron states dominate. One recognizes that
$\rho_{\downarrow}(E)$ has a low-energy tail,  
which covers exactly the same energy region as
$\rho_{\uparrow}(E)$. This tail consists of scattering states  
(magnon emission!) which belong to spin-flip excitations of the
$\downarrow$ electron. Such processes  
are of course possible only when there are $\uparrow$-states within
reach. That explains the coincidence of the low-energy tail of
$\rho_{\downarrow}(E)$  with $\rho_{\uparrow}(E)$ at low temperatures.  
 
With increasing temperature, decreasing moment magnetization 
$\langle S^z\rangle$, more and more magnons are created which can be absorbed by the
$\uparrow$ electron. Consequently, scattering states  
(magnon absorption!) appear in the $\uparrow$ spectrum, too. Q-BS as well
as Q-DOS for $\uparrow$ and $\downarrow$  
continuously approach each other with increasing $T$, at
$T_{\textrm{C}}$ the spin asymmetry is removed.  
However, a correlation-caused splitting of the energy dispersion and a
splitting  of the conduction band into two quasiparticle subbands
remains  which should be observable in a respective photoemission
experiment.
\begin{figure}[ht]
\centerline{\includegraphics[width=0.83\linewidth]{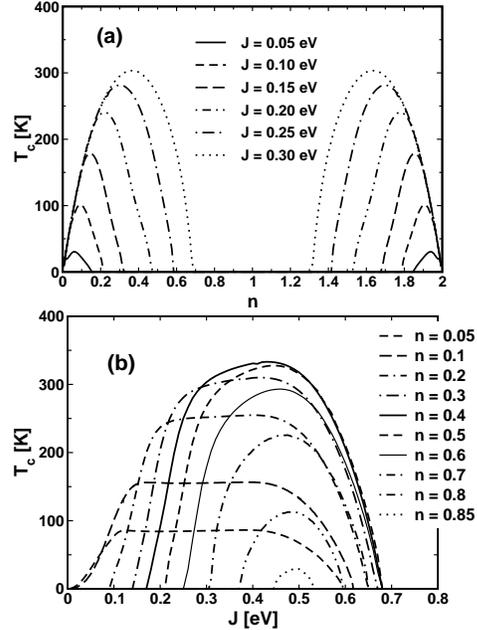}}
\caption{Curie temperature as function of (a) the band occupation $n$, (b)
  the interband exchange coupling $J$ for various (a) $J$, (b) $n$, calculated
  by use of the ''modified'' RKKY \cite{SNO01}. Other parameters as in
  Fig.~\ref{fig:QDOSa}.}
\label{fig:TcnJ}
\end{figure}

The most important entity of ferromagnetism is the Curie temperature
$T_{\textrm{C}}$.   
So it is worthwhile  
to have a look on its dependence on model parameters like band occupation $n$ and exchange  
coupling $J$ (Fig.~\ref{fig:TcnJ}). Low electron (hole) densities appear
to be convenient for a ferromagnetic  
ordering of the local moments, where the ferromagnetic region increases with increasing $J$.  
Around half-filling ($n = 1$) ferromagnetism is excluded. Even more striking and substantially  
deviating from the conventional RKKY picture is the $J$-dependence of
$T_{\textrm{C}}$ (Fig.~\ref{fig:TcnJ}b). Except for  
very low electron densities $n$ there exist a lower and an upper
critical $J_c$. Finite Curie temperatures appear only between these two
limits, which in addition are approaching each other  
with increasing $n$. This is a new feature which goes far beyond
textbook-RKKY and has delicate consequences for the application of the
KLM to strongly coupled ferromagnetic local  
moment systems like Ga$_{1-x}$Mn$_x$As and La$_{1-x}$(Ca, Sr)$_x$
MnO$_3$. On the other hand, the results in  
Fig.~\ref{fig:TcnJ}b do not exclude a reappearance of ferromagnetism for
still stronger $J$. This could not be checked up to now. 
\section{Ferromagnetic Local-Moment Films}\label{sec:films} 
According to the general scheme annotated in the Introduction we are now going to combine  
the preceding model study with a first-principles bandstructure
calculation in order  to investigate the electronic structure of a real
ferromagnetic material. In particular, we want to present  
results for thin ferromagnetic EuO films. Some preparations are still necessary. 
\subsection{Model Films}\label{ssec:modfilm}
We consider a film as a piece of solid consisting of n monolayers parallel to two surfaces. The  
monolayers are numbered by Greek letters ($\alpha,\beta,...= 1,2,..., n$). We assume translational  
symmetry within the two-dimensional surfaces, so that the film can be
described as a two-dimensional Bravais lattice ($\vec{R}_i$) with an $n$-atomic basis ($\vec{r}_{\alpha}$): 
\begin{equation}\label{eq:Ria} 
\vec{R}_{i\alpha} = \vec{R}_i + \vec{r}_{\alpha}                                           
\end{equation} 
The influence of the surface consists in the forbidden electron hopping
in the third 
space-direction and a possibly modified hopping near the surface, which may
give rise to the appearance of surface states. 
 
Because of the two-dimensional translational symmetry the thermodynamic average of any  
site-dependent operator $O_{i\alpha}$ is independent of the
Bravais-index $i$, but may retain a layer-dependence 
$\langle O_{i\alpha}\rangle=\langle O_{\alpha}\rangle$. That holds, in particular, for many-body
terms like Green functions, spectral densities and
Q-DOS. Fourier-transforms use wave-vectors from the  
two-dimensional Brillouin zone of the surface, e.g.: 
\begin{equation}\label{eq:Sigma_ab}
\Sigma^{\alpha\beta}_{\vec{k}\sigma}(E)=\frac{1}{N}\sum_{ij}
\Sigma^{\alpha\beta}_{ij}(E)\:e^{-i\vec{k}\cdot(\vec{R}_i-\vec{R}_j)}
\end{equation}
The many-body concepts, developed for the bulk (Sect.~\ref{sec:KLM}),
remain, however, exactly the same  
for the film. The only difference is that the central equations of the
theory now appear in matrix form. That complicates a little bit the numerical evaluation. 
 
\begin{figure}[ht]
\includegraphics[width=0.485\textwidth]{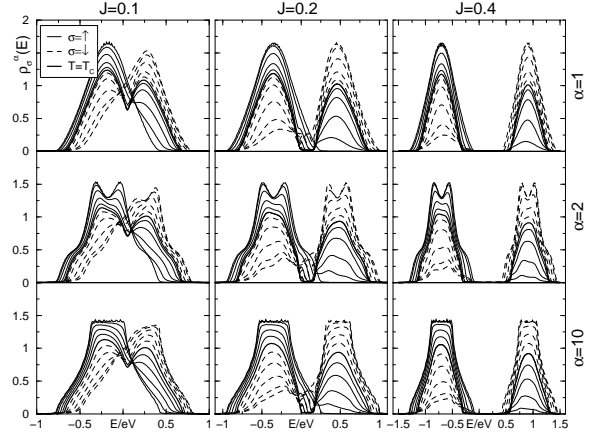}
\caption{Local quasiparticle density of states of the first ($\alpha = 1,20$),
  second ($\alpha = 2,19$),  and center ($\alpha = 10,11$) layer of a 20-layer
  s.c.-(100)  film for different interband
  exchange  couplings $J$ and different temperatures $T/T_{\textrm{C}} = 0, 0.4, 0.7,
  0.9, 0.98, 1$. The bold lines
  are for $T = T_{\textrm{C}}$, the outermost curves belong to $T = 0$ (full lines:
  up-spin,  broken lines: down-spin). Ferromagnetic semiconductor: $n = 0$.}
\label{fig:20-layer}
\end{figure}

The interesting question is whether or not the correlation effects,
worked out for the bulk-KLM in Sect.~\ref{sec:KLM}, are decisively
influenced by the reduced symmetry of a thin film. As an example we
present in Fig.~\ref{fig:20-layer} the local Q-DOS
$\rho^{\alpha}_{\sigma}$ of a semiconducting
20-layer s.c.-film for  
three different exchange couplings $J$ and various temperatures \cite{SNO99}. There
is a clearly visible  
layer dependence of $\rho^{\alpha}_{\sigma}$. However, the physical
interpretation is exactly the same as for the  
bulk (Fig.~\ref{fig:QDOSa}). All quasiparticle features remain valid and
determine the striking temperature  
dependence of the Q-DOS, in film structures, too. A new feature is the possible appearance of  
surface states as a consequence of the modified hopping near and within the surface. That will  
be inspected in Sect.~\ref{ssec:EuOfilm} with respect to a EuO (100)-film. 
\subsection{Multiband Kondo Lattice Model}\label{ssec:MKLM}
To apply our model study to a ferromagnetic EuO (100) film we have first to remove some  
model simplifications. The (empty) conduction bands of EuO have $5d$ character. So we have  
to replace the previous assumption of a non-degenerate $s$-band
accordingly. Instead of (\ref{eq:hamil}) we write 
\begin{equation}\label{eq:MKLM}
H=H_d+H_f+H_I                                                
\end{equation}
$H_d$ refers to the conduction bands: 
\begin{equation}\label{eq:Hd} 
H_d=\sum_{ij\alpha\beta\sigma}\sum_{mm^{\prime}}T^{mm^{\prime}}_{ij\alpha\beta}
c^{\dagger}_{i\alpha m\sigma}c^{}_{j\beta m^{\prime}\sigma}                                 
\end{equation}
$m$, $m^{\prime}$ denote different
orbitals. $T^{mm^{\prime}}_{ij\alpha\beta}$ are the hopping integrals
($\vec{R}_{i\alpha} \leftrightarrow \vec{R}_{j\beta}$) which we take  
from a first principles bandstructure calculation according to the tight-binding LMTO-ASA  
program of Anderson \cite{AND75,AJE84}. The conduction band of the semiconductor
EuO is empty at $T=0$. From the exact limiting case of the KLM
(Sect.~\ref{ssec:limcase}) we know that in ferromagnetic saturation the
$\uparrow$  spectrum is identical to the ''free'' Bloch-spectrum except
for an unimportant rigid shift (Fig.~\ref{fig:QDOS}, Fig.~\ref{fig:QDOSa}). So we take
from the bandstructure calculation the $\uparrow$ spectrum as input for  
$H_d$, thereby elegantly circumventing the double counting of the interband exchange. We do  
not avoid the double counting by ''switching off'' the relevant interaction part, what appears  
almost impossible, but by exploiting the exact limiting case, for which this interaction part  
causes only a rigid shift of the $\uparrow$ spectrum. Remember that the
$\downarrow$ spectrum, on the contrary, is  
influenced by the interband exchange in a drastic manner, even at $T = 0$. 
 
$H_f$ refers to the local moment system (half-filled $4f$ shell of the
Eu$^{2+}$ ion), which  we describe by an extended Heisenberg Hamiltonian: 
\begin{equation}\label{eq:extHeis}
H_f=-\sum_{ij\alpha\beta}J^{\alpha\beta}_{ij}\:\vec{S}_{i\alpha}
\cdot\vec{S}_{j\beta}-D_0\:\sum_{i\alpha}\left( S^{z}_{i\alpha}\right)^2
\end{equation}
The first term is the original Heisenberg Hamiltonian (\ref{eq:Hf}),
while the second term introduces a  
symmetry-breaking single-ion anisotropy being necessary to overcome the Mermin-Wagner  
theorem which excludes a collective magnetic order in the isotropic Heisenberg model for  
film geometries at finite temperatures \cite{MWA66,GNO00,GNO01}. Because of the empty
conduction bands a self-consistent justification of the
EuO-ferromagnetism via an RKKY-type mechanism is not possible. Instead
of this we take the experimental values for the exchange integrals   between  
nearest ($J_1$) and next-nearest neighbours ($J_2$)\cite{BZD80} 
($J_1/k_{\textrm{B}} = 0.625K; J_2/k_{\textrm{B}} = 0.125K$). 
 
The third term in (\ref{eq:MKLM}) describes the on-site Coulomb
interaction between electrons in different  
orbitals. By a straightforward consideration\cite{SMN01} one finds an
interaction  operator in strict generalization of (\ref{eq:Hsf}): 
\begin{equation}\label{eq:HiMB}
H_I=-J\sum_{i\alpha m}\boldsymbol{\sigma}_{i\alpha m}
\cdot\vec{S}_{i\alpha}
\end{equation}
Although the multiorbital situation complicates once more the numerics of the evaluation,  
nevertheless one can apply the same many-body concepts, successful for the simple KLM  
(Sect.~\ref{sec:films}), to the multiband-KLM (\ref{eq:MKLM}), too. In the next Section a short
list of some typical results is presented. 
\subsection{EuO (100) Films}\label{ssec:EuOfilm} 
\begin{figure}[ht]
\includegraphics[width=0.4\textwidth]{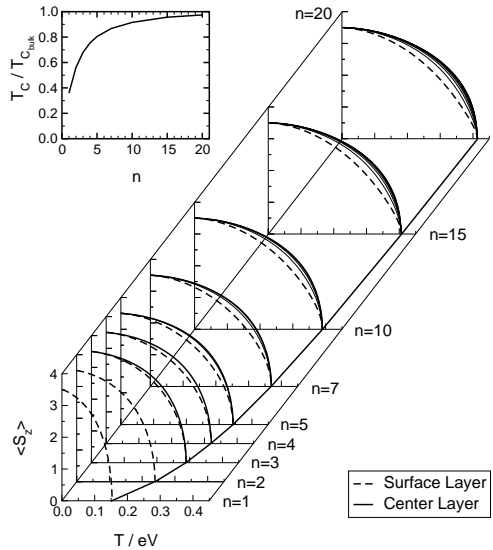}
\caption{Layer-dependent magnetizations of EuO(100) films as a function
  of  temperature and for various film thicknesses ($n$ = number of
  monolayers). Parameters $J_1/k_{\textrm{B}} = 0.125 K$, $J_2/k_{\textrm{B}} =
    0.125K$,  $D_0/k_{\textrm{B}}= 0.05 K$.  Inset: Curie temperature
    as function of film thickness ($T^{\textrm{Bulk}}_{\textrm{C}}=69.33K$).}
\label{fig:temp-layer}
\end{figure}

Fig.~\ref{fig:temp-layer} shows the temperature- and layer-dependent
magnetization  for films of different  
thickness (from $n = 1$ to $n = 20$ monolayers). For all temperatures and all thicknesses    
the magnetization $\langle S^z_{\alpha}\rangle$ increases from the
surfaces ($\alpha = 1, n$) to the center layers
($\alpha =\frac{n}{2}$ or $\frac{n+1}{2}$), qualitatively understandable
because of the reduced coordination number of the surface atoms. The
Curie temperature $T_{\textrm{C}}$ is obviously strongly
thickness-dependent. Starting from about $15K$ for the  
monolayer $T_{\textrm{C}}$ steadily increases with $n$ reaching the bulk
value ($69.33K$ \cite{WAC79}) for $n\approx 30$. The  
$20$-layer film has $T_{\textrm{C}} = 66.7K$. The $T_{\textrm{C}}(n)$
curve in Fig.~\ref{fig:temp-layer} remarkably resembles that of Gd films
\cite{FBS93}. To our knowledge corresponding measurements on EuO films have not been done yet.  
 
\begin{figure}[ht]
\includegraphics[width=0.45\textwidth]{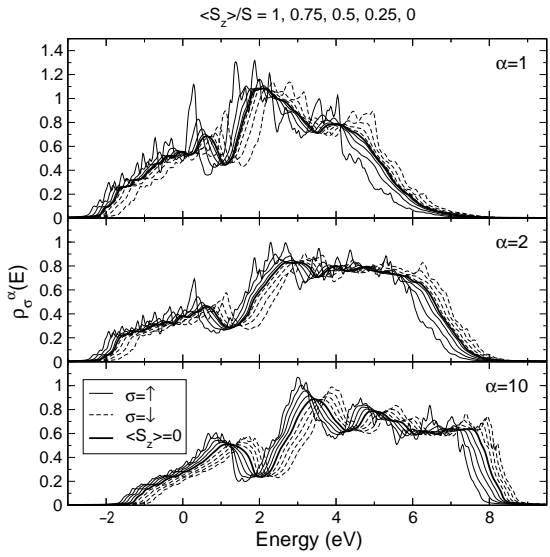}
\caption{Local quasiparticle densities of states of the empty Eu-$5d$
  bands  of the first ($\alpha = 1$), second ($\alpha = 2$) , and center ($\alpha = 10$)
  layers of a 20-layer EuO(100) film for different temperatures ($T_{\textrm{C}} = 66.7K$).}
\label{fig:QDOS20layer}
\end{figure}

The temperature-dependence of the local moment correlations $\langle
S^z_{\alpha}\rangle$,$\langle
S^{\pm}_{\alpha}S^{\mp}_{\alpha}\rangle$,\dots transfers via  
the interband coupling $J$ to the (empty) band states (Fig.~\ref{fig:QDOS20layer}). The
shift  of the layer-dependent Q-DOS with temperature is not at all rigid
but with strong irregularities, 
mainly due to the above-discussed spinflip
processes. Fig.~\ref{fig:QDOS20layer} shows $\rho^{\alpha}_{\sigma}(E)$ for a 20-layer film.  
The lower $\uparrow$-band edge shifts upon cooling from $T =
T_{\textrm{C}}$ down to $T = 0K$ by some $0.3eV$. This  
is the famous ''red shift'', experimentally observed as corresponding
shift of the optical absorption edge already some 35 years ago \cite{WAC79}. 
 
The Q-DOS of the center layer ($\alpha = 10$) resembles already pretty
well  the Q-DOS of bulk-EuO \cite{SNO01a}. Comparing it with the surface Q-DOS
($\alpha = 1$) one observes a shift of $\rho^{\alpha=1}_{\sigma}(E)$ to  
somewhat lower energies. This is an indication for the appearance of a
surface  state. A surface state is a state which exists in the so-called
forbidden region where no bulk states occur. Typically the spectral
weight of a surface state decreases exponentially with increasing distance  
from the  
\begin{figure}[ht]
\centerline{\includegraphics[width=0.48\textwidth]{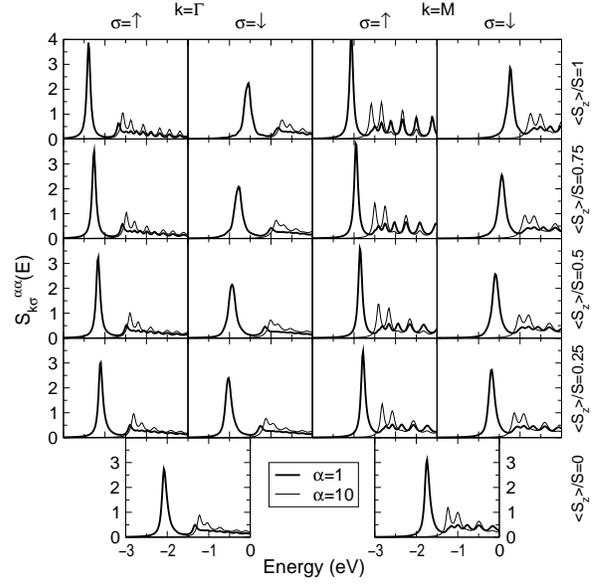}}
\caption{Layer-dependent spectral density of the surface ($\alpha = 1$:
  thick lines) and the center layer ($\alpha = 10$: thin lines) of a 20-layer
  EuO(100) film at the $\Gamma$ point and the $\textrm{M}$
  point  for $J = 0.25 eV$ and for different temperatures ($T_{\textrm{C}} = 66.7K$).}
\label{fig:QSD20layer}
\end{figure}
surface. Fig.~\ref{fig:QSD20layer} proves the existence of the EuO(100)
surface  state by the spectral density $S^{\alpha\alpha}_{\vec{k}\sigma}(E)$
of the 20-layer film at $\vec{k}=\bar{\Gamma}$  and
$\vec{k}=\bar{\textrm{M}}$. Below the $\alpha = 10$ spectrum there is for  
both $\vec{k}$-values a prominent peak of the spectral density for 
$\alpha =1$. For $T = 0$ the surface state  
lies at the $\bar{\Gamma}$ point about $0.8eV$ and the
$\bar{\textrm{M}}$  point about $0.45eV$ below the ''bulk'' ($\alpha = 10$)  
spectrum. These splittings between surface states and the lower edges of
the ''bulk'' spectrum  are practically temperature-independent. The
induced spin splitting of the surface state as well  
as that of the band states collapses with increasing temperature 
$T \longrightarrow T_{\textrm{C}}$. 
 
This temperature-behaviour gives rise to an interesting speculation
\cite{SNO01b}. The gap between the  
occupied $4f\uparrow$-states and the empty $5d$-conduction band states
amounts to $1.12eV$  at $T = T_{\textrm{C}}$ \cite{WAC79}.  
The lower edge of the surface band lies $0.8eV$ below the $5d$-edge. The red shift of about  
$0.3eV$ will further reduce the gap when decreasing the temperature to
$T = 0$. The overall gap reduction ($\approx 1.1eV$) is therefore in the range of
the experimental $4f-5d_{t_{2g}}$ gap of bulk EuO at  
room temperature. That makes a surface insulator-metal transition in EuO(100) films possible when the  
film is cooled down to $T\longrightarrow 0K$. Because of the induced
exchange splitting of the surface  
states $\uparrow$ electrons will tunnel from the $4f$ band into the
surface band  resulting in a halfmetal.  
Furthermore, the resistivity of the EuO(100) film should be highly reactive to an external  
magnetic field giving rise to a colossal magnetoresistance effect. 
\section{Conclusions }\label{sec:conc}
With the concrete example of a thin ferromagnetic EuO(100) film we have demonstrated our  
method of determining magnetic properties of real materials as
consequencies of the  temperature-dependent electronic structure. The
method combines the many-body evaluation of a  
properly chosen theoretical model with a first principles bandstructure calculation in order to  
get more or less quantitative electronic structure information. 
 
EuO belongs to the so-called local-moment ferromagnets which are reasonably modelled by  
the Kondo lattice model. We have therefore inspected in detail the properties of this model,  
first for the bulk and then for film geometries. The results have been brought into contact with  
a tight-binding linear muffin-tin orbital bandstructure calculation (TB-LMTO). The resulting  
temperature-dependent electronic structure gave rise to the speculation
of a surface insulator-halfmetal transition when cooling the film below $T_{\textrm{C}}$. The
reason is the Stoner-like temperature  
shift of an empty $5d$ surface state (band) with an induced exchange
splitting. A colossal magnetoresistance effect may be expected. 
\section*{Acknowledgements}\label{sec:ack}
A great part of this review was fuelled by the PhD.~thesis of R.~Schiller, submitted at the  
Humboldt-Universit{\"a}t zu Berlin (Germany). Financial support by the
SFB~290  of the ''Deutsche Forschungsgemeinschaft'' and by the
''Volkswagen-Stiftung'' is gratefully acknowledged.

\end{document}